\documentclass{ws-book9x6}
\usepackage{ws-book-thm}   
\usepackage{ws-book-har}   
\usepackage[pdfpagelabels=false,colorlinks=true,allcolors=black]{hyperref}  

\makeindex

\begin{document}
\chapter{Dynamical Heterogeneity in Glass-Forming Liquids}\label{ch1}
\author{Giulio Biroli}
\address{Laboratoire de Physique de l'Ecole Normale Sup{\'e}rieure, ENS,
Universit\'e PSL, CNRS, Sorbonne Universit\'e, Universit\'e de Paris, F-75005 Paris, France} 
\author{Kunimasa Miyazaki}
\address{Department of Physics, Nagoya University, Nagoya 464-8602, Japan}
\author{David R.~Reichman}
\address{Department of Chemistry, Columbia University, 3000 Broadway, New York, New York 10027, USA, USA}

\section{Introduction: dynamical correlations in glassy dynamics}
\begin{figure}[ht]
\centerline{
\includegraphics[width=3.9cm]{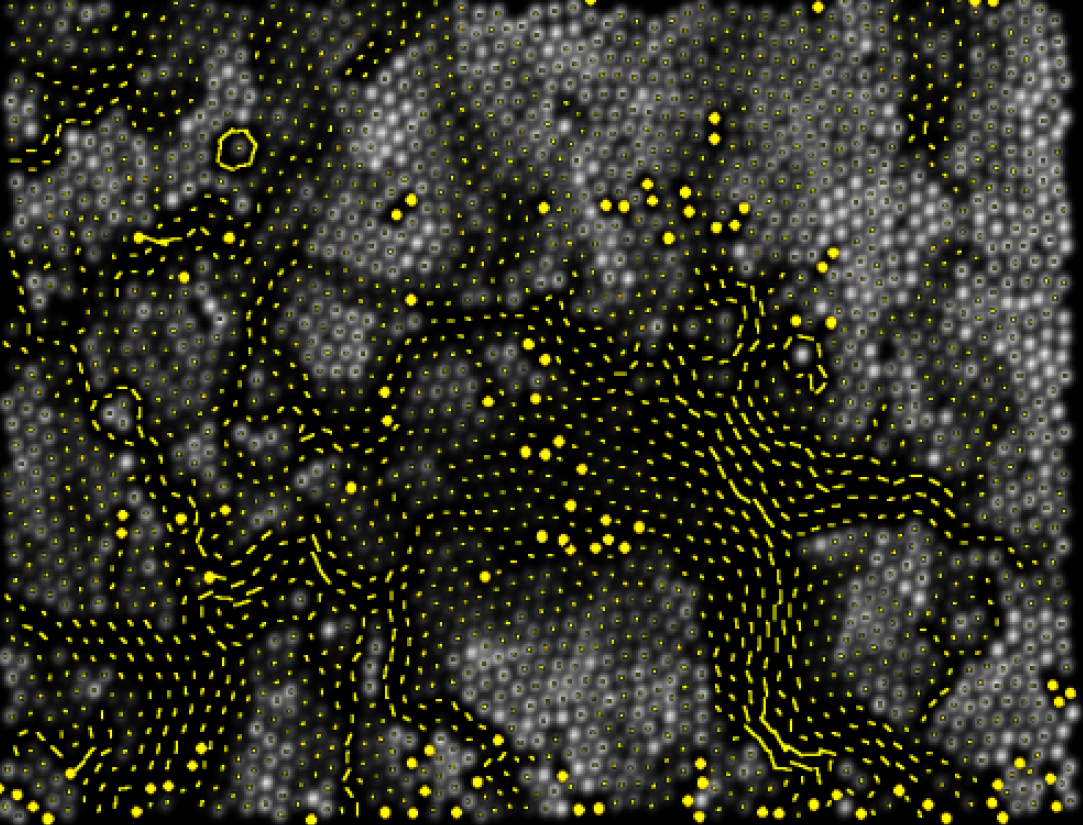}
\includegraphics[width=3.9cm]{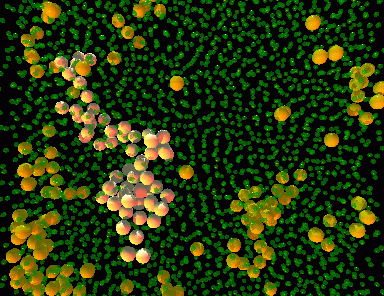}
\includegraphics[height =3cm]{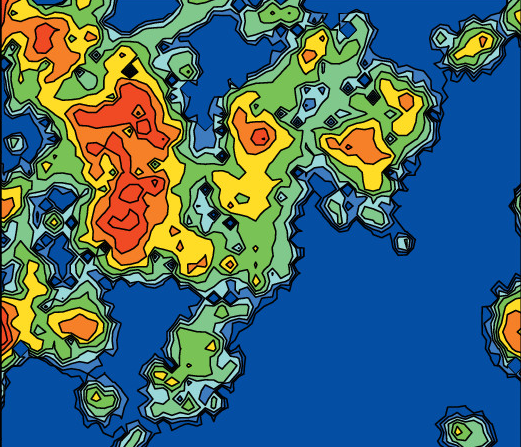}}
\caption{Dynamical heterogeneity in a granular fluid of metallic discs (left)~\cite{Dauchot2005prl},
 in a colloidal hard sphere suspension (centre)~(Courtesy of Eric R. Weeks and David A. Weitz, previously unpublished), 
and in computer simulated two-dimensional system of repulsive disks~\cite{WidmerCooper2009jcp}.
In all cases, clusters of high and low mobility are highlighted.}  
\label{fig1}
\end{figure}

A well-known puzzle for researchers working on the glass transition problem is that a static snapshot of a supercooled liquid looks very similar to that of a high-temperature liquid. However, these two systems are very different {\it dynamically}, as their relaxation times (henceforth denoted $\tau_\alpha$) can differ by fourteen orders of magnitude\footnote{
The relaxation time is usually defined as the time $t$ at which a finite
fraction of the system, say one half, is relaxed.  
In practice, $\tau_\alpha$ is the time $t$ at which a suitable global time-dependent correlation function has decreased by, say, one-half.}. No simple signature of this phenomenon is found in particle configurations\footnote{The understanding of this state of affairs has changed in recent years through the discovery of subtle static correlations, see \cite{berthier2011theoretical}.}. What has been understood in the last thirty years is that a clear signature can instead be found by looking at dynamical correlations or, expressed differently, by observing how the dynamical relaxation process unfolds in space and time.

Dynamical relaxation events are correlated in space, and these spatial correlations grow approaching the glass transition \cite{berthier2011dynamical}. In order to understand this phenomenon, one must focus on  a \emph{mobility field} which measures the instantaneous relaxation that has taken place in a window of time $t$ at a given position ${\bf r}$~\cite{kob1997dynamical}
\begin{equation}\label{eq:defc}
    c({\bf r} ; t,0) = \sum_i c_i(t,0) \delta({\bf r}-{\bf r}_i),
\end{equation}
where the sum is over all particles (indexed by $i$) and the mobility field $c_i(t,0)$ is a two-point function which compares the configuration at time $0$ with that at time $t$. 
For example, to measure relaxation on a length scale $2 \pi / q$, one might consider $o_i(q,t)=e^{i{\bf q}\cdot {\bf r}_i(t)}$
and $c_i(t,0)=o_i({\bf q},t) o_i(-{\bf q},0)$.  In this case,
$o_i({\bf q},t)$ is related to a Fourier component of the density
of the system, and the average of $c_i(t,0)$ is the self-part
of the so-called intermediate scattering function $F({\bf q},t)$ \cite{hansen1988theory}. Many other choices have also been used in the literature (see Ref.~\cite{berthier2011dynamical} for other examples). 

Contrary to its static counterpart, the statistics of the mobility field does show remarkable changes upon supercooling. By considering windows of time of the order of $\tau_{\alpha}$, one finds that dynamics becomes progressively more heterogeneous and correlated in space. We show in Fig.~\ref{fig1} three examples of dynamical heterogeneity for three different kinds of glassy liquids: a granular fluid, a colloidal suspension, and an atomistic simulation. In all cases, one finds that relaxation emerges in correlated clusters of size $\xi(t)$. By studying the $t$-dependence of dynamical heterogeneity, one finds that such clusters, and accordingly $\xi(t)$,  increase with time until occupying a substantial fraction of the system size for $t=\tau_\alpha$.

\begin{figure}[ht]
\centerline{\includegraphics[width=8cm]{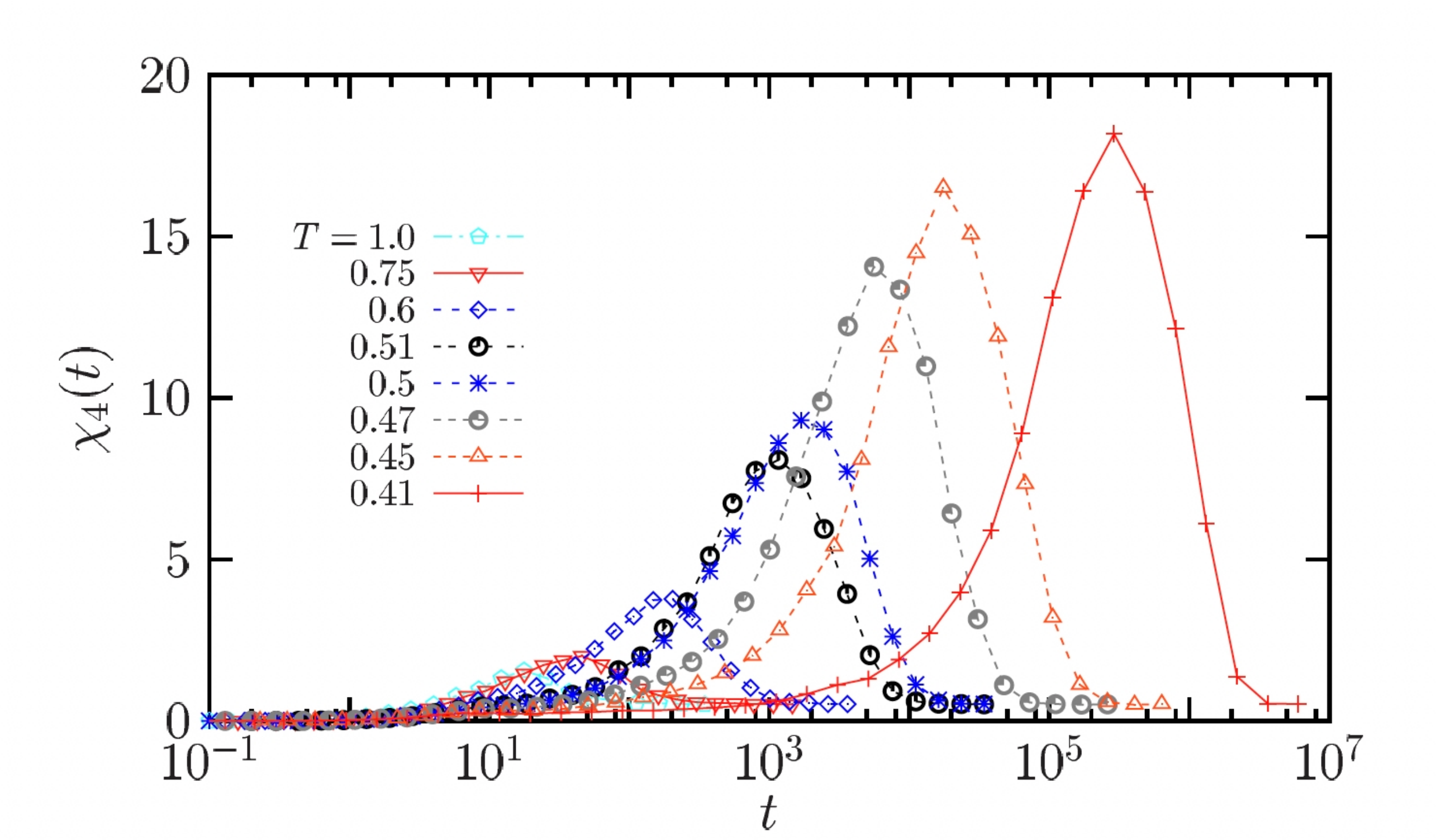}}
\caption{Time dependence of $\chi_4(t)$ obtained using the 
self-intermediate scattering 
function for $C(t,0)$ from a molecular dynamics simulation of a 
Lennard-Jones supercooled liquid.
}
\label{fig2}
\end{figure}

In order to probe and measure the spatial correlations of the mobility field, two important kinds of correlation functions have been introduced. The first is the 
the spatial correlation function~\cite{dasgupta1991there}
\begin{equation}
    G_4(r;t) = \langle c({\bf r};t,0) c({\bf 0};t,0) 
\rangle - \langle c({\bf r};t,0) \rangle^2,
\label{eq:g4def}
\end{equation}
where brackets denote thermal average over equilibrium dynamics.  This expression is the equivalent to the two-point correlation function in critical phenomena. By analyzing its dependence (decrease) on ${\bf r}$ one can obtain $\xi(t)$. The second function to have played a very important role in the study of glassy dynamics is the analog of the susceptibility in ordinary phase transitions. Such a function is defined in terms of the fluctuations of a {\it global} time-dependent correlation function $C(t,0)$ \cite{kob1997dynamical}
\begin{equation}
    \chi_4(t) = N [ \langle C(t,0)^2 \rangle - \langle C(t,0) \rangle^2 ],
\label{eq:chi4var}
\end{equation}
where $N$ is the number of particles in the system. 
If one thinks of $C(t,0)$ as the order parameter of the glass transition, $\chi_4(t)$ measures its fluctuations (see Sec.~\ref{sec:chi4} for more details). 
Note that since $C(t,0)$ is itself a two-point correlation function, as it compares the system at two different times, the function $\chi_4(t)$ is a four-point function. Again, as in critical phenomena, one can link the two kinds of correlation functions as 
\begin{equation}
    \chi_4(t) = \int\mathrm{d}r\, G_4(r;t).
\end{equation}
One therefore expects that if 
\begin{equation}
    G_4(r; t) \sim \frac{A(t)}{r^{p}} e^{-r/\xi_4(t)}
\end{equation}
with $p$ a suitable exponent then $\chi_4(t)$ measures the typical number of particles
involved in correlated motion (assuming that the prefactor $A(t)$ does not change considerably with $t$). We show in Fig.~\ref{fig2} the behavior of $\chi_4(t)$
in an atomistic simulation for different degrees of supercooling. One indeed finds that $\chi_4(t)$ increases as a function of time at fixed temperature, and that its peak increases when decreasing the temperature.

The realization that glassy dynamics becomes progressively more correlated in space--thanks to the introduction of the above correlation functions--has opened the way to a field-theoretical analysis of the glass transition and to strong connections with replica theory. The aim of this chapter is to briefly review this relationship, the impact that ideas and methods from replica theory have had on the field of dynamical correlations, and discuss future research directions. 

Before concluding this introduction, we wish to stress that the dynamical heterogeneity of glassy dynamics encompasses a variety of phenomena that is broader than the one we briefly recalled above \cite{berthier2011dynamical}. 
The existence of dynamical correlations is strongly related to the existence of dynamical facilitation; assessing how facilitation influences the space-time correlations associated with dynamical heterogeneity has been a very important topic in the field \cite{biroli2013perspective}. Furthermore, the realization that to characterize glassy dynamics one has to focus on high-order dynamical correlation functions has suggested that likewise non-linear response functions should play an important role \cite{bouchaud2005nonlinear}. This insight has been confirmed, and is now used to characterize spatial properties of glassy dynamics in experiments \cite{albert2016fifth}. We refer to the book \cite{berthier2011dynamical} for a more thorough introduction to the field of dynamical heterogeneity. We will come back to some of the points discussed above at the end of this chapter when discussing future research directions.    

\section{$p$-spin model and dynamical heterogeneity}
\label{sec:pspin}
The theoretical underpinnings of dynamical heterogeneity in supercooled liquids and glasses owe much to simplified models.  Perhaps no model has had a larger impact on the study of dynamical heterogeneity, or the glass transition as a whole, than the $p$-spin spin glass model. In this section, we go over the key physical highlights of that model.
\subsection{RSB and metastable states}
The $p$-spin model was first introduced in the context of the glass transition by Kirkpatrick and Thirumalai in 1987 as an exactly solvable toy model whose dynamical solution coincides precisely with that of the schematic version of the microscopic mode-coupling theory (MCT) of liquid-state dynamics~\cite{KT87}.  Specifically, starting from the  Hamiltonian
\begin{equation}
    H=-\sum_{1 \leq i_{1}<i_{2} \cdots <i_{p}\leq N }J_{i_{1} i_{2} \cdots i_{p}}s_{i_{1}}s_{i_{2}}\cdots s_{i_{p}}
\end{equation}
with integer $p$ and $p>2$, Gaussian couplings with zero mean and variance $\frac{p!}{2N^{p-1}}$, and the constraint $\frac{1}{N}\sum_{i=1}^{N}s^{2}_{i}=1$, one finds the exact dynamical equation for the spin-spin correlation function $C(t)=\frac{1}{N}\sum_{i=1}^{N}\langle s_{i}(t)s_{i}(0)\rangle$ (brackets denote average over thermal noise and quenched couplings)
\begin{equation}
    \frac{\partial C(t)}{\partial t}+TC(t)+\frac{p}{2T}\int_{0}^{t}d\tau C^{p-1}(t-\tau)\frac{\partial C(\tau)}{\partial \tau}=0
\label{eq:eom}
\end{equation}
with $C(0)=1$ and $T$ the temperature set by the thermal noise. This non-linear integro-differential equation is identical to the schematic equation of motion for the density correlator for $p=3$~\cite{Gotze,CharbReich}.  Although not specifically stated in the original work of Kirkpatrick and Thirumalai, the construction of such an exact equation from an underlying model follows the same logical path taken by Kraichnan decades before in the formulation of the direct interaction approximation (DIA) for the Navier-Stokes equation~\cite{DIA}.  In particular, the introduction of quenched randomness exactly renders all diagrams from the field-theoretic solution subleading except for the \emph{melonic} diagrams~\cite{BCKM}.  In this sense, the original work of Kirkpatrick and Thirumalai amounts to a realizable model for the MCT equations of liquids.  The very same considerations are at play in the exact solution to quantum models such as the SYK model~\cite{SYK}.  The realizability of the $p$-spin model means that one can take the model more literally, and explore its properties beyond those afforded by the already known high temperature properties of the schematic MCT equations.  These more global properties form a mean-field foundation for the random first-order theory (RFOT) of the glass transition as formulated by Kirkpatrick, Thirumalai and Wolynes two years later~\cite{KTW}.

The dynamical properties of the model for $p>2$ can be divided into two regimes.  In the high temperature regime the fluctuation-dissipation theorem (FDT) holds and only the spin-spin correlator, which is a function of time differences only, is required for a complete solution.  This quantity decays as a single exponential function in time if temperature is sufficiently high. As the temperature is lowered, a plateau in the relaxation appears. The duration of the plateau grows in time as the temperature is lowered until a sharp transition to a non-ergodic behavior at a temperature $T_{d}$ occurs.  This transition to an arrested state can be viewed as a purely dynamical phenomenon, although as we will discuss below, the transition can be given a thermodynamic-like meaning. Below $T_{d}$ the FDT is violated and aging behavior sets in.  This behavior, which requires the consideration of both the spin-spin correlation function and the associated response function unlike the simpler MCT equation written above, was first exactly solved by Cugliandolo and Kurchan in 1993~\cite{CK}.

The replica method provides an essentially complete thermodynamic
picture of the free energy of the $p$-spin model at all temperatures~\cite{Cavagna,Parisi}. The high temperature regime above $T_{d}$ corresponds to a replica symmetric solution where the free energy is smooth and contains only one basin corresponding to the ergodic \emph{liquid} state.  The transition at $T_{d}$ is a harbinger of replica symmetry breaking which occurs in one step (1RSB, by contrast to the infinite step, $\infty$RSB, that occurs in the Sherrington-Kirkpatrick model~\cite{Parisi}).  Below $T_{d}$ the free energy fractures into an extensive number of metastable free energy minima separated by barriers that are infinitely large in the thermodynamic, $N \rightarrow \infty$, limit.  The configurational entropy, or complexity, counts the number of metastable states in a given energy range.  At a temperature $T_{K}<T_{d}$, the complexity becomes subextensive in $N$.  This \emph{entropy vanishing} transition, which occurs deep in the glass state, may be viewed as the analog of the empirically-defined Kauzmann transition, where an entropy crisis (the crossing of the configurational entropy associated with the crystal and glass) is envisioned to occur in real materials~\cite{Debenedetti}.  It should be noted that in some variants of $p$-spin and related models, such as the hard-spin (Ising) version of the $p$-spin model, $\infty$RSB may occur where the free energy landscape takes on a hierarchical structure~\cite{Gardner,Gross}.  Evidence for this type of transition (the Gardner transition) also appears to find some support in more realistic off-lattice simulation models, although we will not discuss this behavior further~\cite{LudoGardner}.

 \subsection{$\chi_4(t)$ and the dynamical overlap} 
 \label{sec:chi4}
The dynamical behavior of the $p$-spin model is also quite remarkable. One can define a static overlap $Q=\frac{1}{N}\sum_{\alpha}s_{\alpha}s{'}_{\alpha}$ which measures how similar two different configurations denoted by $\bf{s}$ and $\bf{s}{'}$ are.  The logarithm of the probability distribution of the static overlap defines an effective potential, called the Franz-Parisi potential, which exhibits non-trivial features as temperature is lowered~\cite{FP}.  In particular, at high temperatures the Franz-Parisi potential exhibits a single minimum centered at $Q=0$, indicating that the stable phase of the model is a completely disordered \emph{liquid} phase. As temperature is lowered, the function begins to lose convexity, eventually developing a second minimum away from $Q=0$ below $T_{d}$.  Within mean-field theory this static behavior has important implications for the dynamics.  In particular, it implies that there is a diverging dynamical length scale upon approaching the dynamical transition which is accompanied by dynamically heterogeneous behavior.  

To quantify and characterize dynamical heterogeneity, and by analogy the notion of a diverging dynamical length scale, we can generalize the definition of the overlap to consider configurations at different times, $Q(t)=\frac{1}{N}\sum_{i}s_{i}(0)s_{i}(t)$.  The measure of fluctuations of this quantity, $\chi_{4}(t)=N (\langle Q(t)^{2}\rangle-\langle Q(t)\rangle^{2})$, is the precise analog of the function $\chi_{4}(t)$ defined in Eq.~\eqref{eq:chi4var}.  If the system has a dynamical critical point with a diverging length scale, then $\chi_{4}(t)$ should diverge as $T \rightarrow T_{d}$ from above. Indeed, this is precisely what happens in the $p$-spin model.  Because the model has no spatial scale, any analog of $G_{4}(r;t)$ is not meaningful, and the diverging length scale must be inferred from the behavior of $\chi_{4}(t)$ itself.  In a physical sense the model is however clearly heterogeneous in the following manner: for a given realization of disorder, the behavior of the local spin-spin correlation function varies from site to site.  When averaged over all sites and disorder realizations, the variance of these local dynamical fluctuations diverges at the dynamical critical point.

\begin{figure}[ht]
\centerline{\includegraphics[width=8cm]{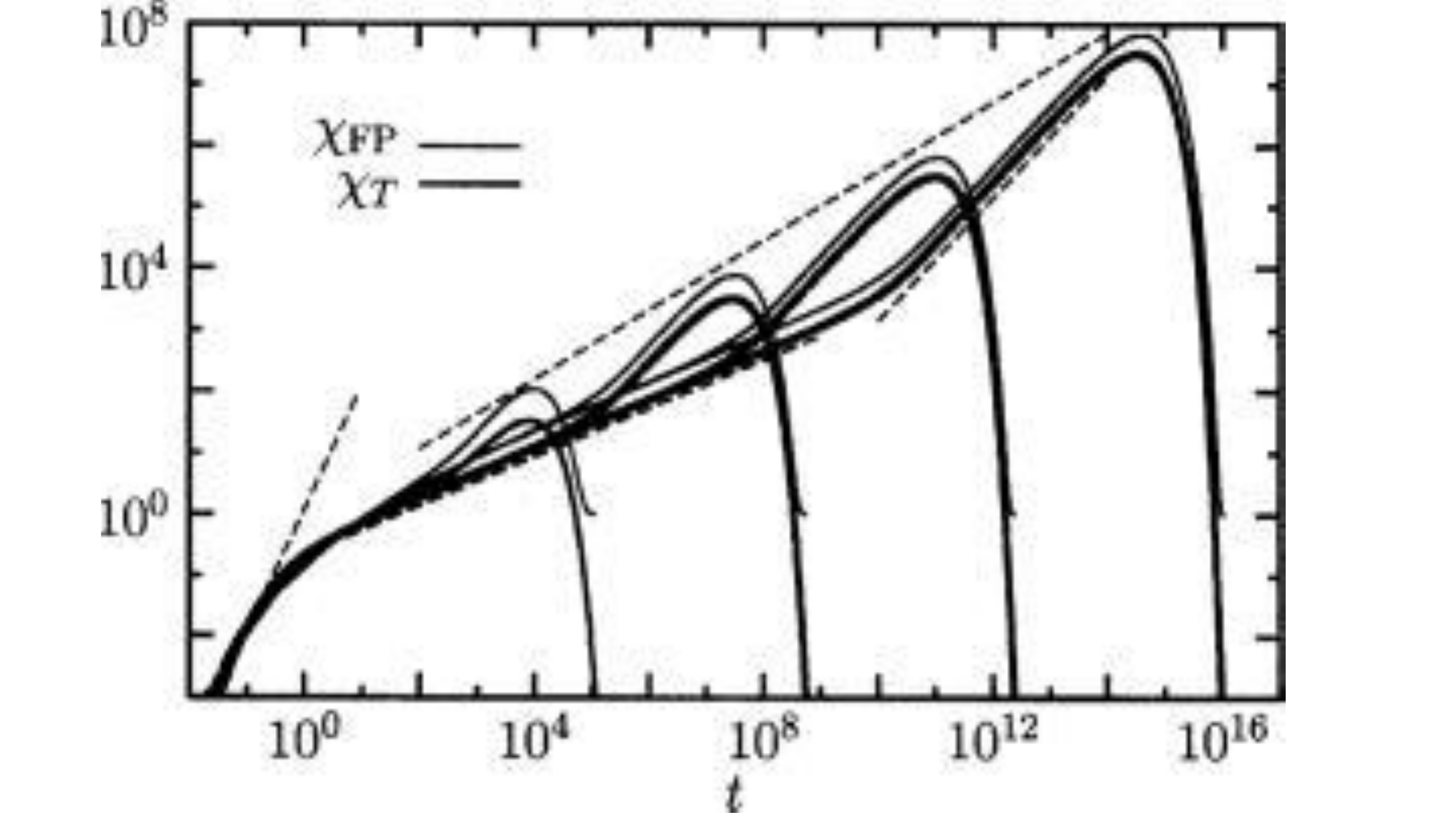}}
\caption{Growth of $\chi_{FP}(t) \sim \chi_{4}(t)$ in the $p$-spin model for $p$=3 as the dynamical transition $T_{d}=0.612$ is approached from above.  Comparison is made to $\chi_{T}=\frac{dC(t)}{dT}$, which is discussed in Ref.~\cite{Berthier2007jcp2}.  Adapted from Ref.~\cite{Berthier2007jcp2}}
\label{fig3}
\end{figure}

The arguments leading to the formulation of $\chi_{4}(t)$ for the $p$-spin model and the calculation of its growth as $T_{d}$ is approached were first put forward by Franz and Parisi in 2000~\cite{FP2000}.  Technically, Franz and Parisi defined a closely related function, $\chi_{FP}(t)$, which is simpler to calculate directly in the $p$-spin model, and is closely related to definition of $\chi_{\bf{k}}({\bf q},t)$ within the inhomogeneous MCT (IMCT) formulation discussed in Sec.~\ref{sec:imct}. This work was influential in motivating the first calculation of $\chi_{4}(t)$ in molecular dynamics simulations 
of supercooled liquids by Glotzer and coworkers~\cite{Glotzer}.  It should be noted that a decade prior to the work of Franz and Parisi, Kirkpatrick and Thirumalai outlined the behavior of dynamical overlap fluctuations in Potts glasses, which are in the same 1RSB class as the $p$-spin model~\cite{KT1988}.  Kirkpatrick and Thirumalai calculated the behavior of the four-point correlator via the summation of ladder diagrams, noting that the ladder sum diverges at $T_{d}$, implying a diverging dynamical length scale at the transition in the model.  It is interesting that this mode of calculation is distinct from the approach taken by Franz and Parisi, yet leads to identical conclusions.  The summation of ladder diagrams was employed by Bouchaud and Biroli in the first attempt at formulating a microscopic liquid-state theory for $\chi_{4}(t)$~\cite{BB2004}.  Lastly, it has been argued in ~\cite{Berthier2007jcp1} that the behavior of $\chi_{FP}(t) \sim \chi_{4}(t)$ should be nearly identical to that of the simpler quantity $\chi_{T}=\frac{dC(t)}{dT}$ (see Sec.~\ref{sec:imct}). The behavior of the latter two quantities are illustrated in Fig.~\ref{fig3}

It is natural to compare the detailed scaling behavior that emerges from
the study of dynamical heterogeneity and $\chi_{4}(t)$ in the $p$-spin
model to that found in {\em in silico} studies of supercooled
liquids~\cite{berthier2004time}.  The scaling behavior in the $p$-spin
model and in liquid-state MCT, where the spatial structure of the
associated features of dynamical heterogeneity is more
explicit~\cite{Biroli2006prl}, is subtle, and will be explicated in
detail below.  On a qualitative level, examination of the simulated
growth of $\chi_{4}(t)$ in the $p$-spin model as illustrated in
Fig.~\ref{fig3} is quite similar to that seen in computer simulations as
shown already in Fig.~\ref{fig2}.  However, on a quantitative level such
mean-field approaches do not capture the space-time scaling properties
associated with dynamical heterogeneity.  The discrepancies between the
predicted mean-field behavior and those extracted from liquid-state
simulations require care to discern.  Using simulations up to $N=10,000$
particles, Stein and Andersen found scaling exponents in quantitative
agreement with those presented in the next section~\cite{Stein}.
However, more extensive simulations by Karmakar {\it et al.} with up to
$N=300,000$ find that, for example, the peak of $\chi_{4}(t)$ and the
dynamical correlation length grow with exponents that differ from those
predicted from the IMCT discussed in
Sec.~\ref{sec:imct}~\cite{Karmakar}.  Given the fact that in three
dimensions, mean-field behavior is modified by effects such as local
particle hopping and dynamical
facilitation~\cite{berthier2011theoretical}, these discrepancies are not
surprising, and in fact the qualitative agreement between particle-based
simulations and mean-field theory provides at least some evidence  
that the theory provides a reasonable picture of supercooled behavior and a foundation for a more developed understanding of glassy behavior in low spatial dimensions.

\subsection{Dynamical heterogeneity and MCT}
\label{sec:imct}
\begin{figure}[t]
\begin{center}
\includegraphics[scale=0.4,angle=0]{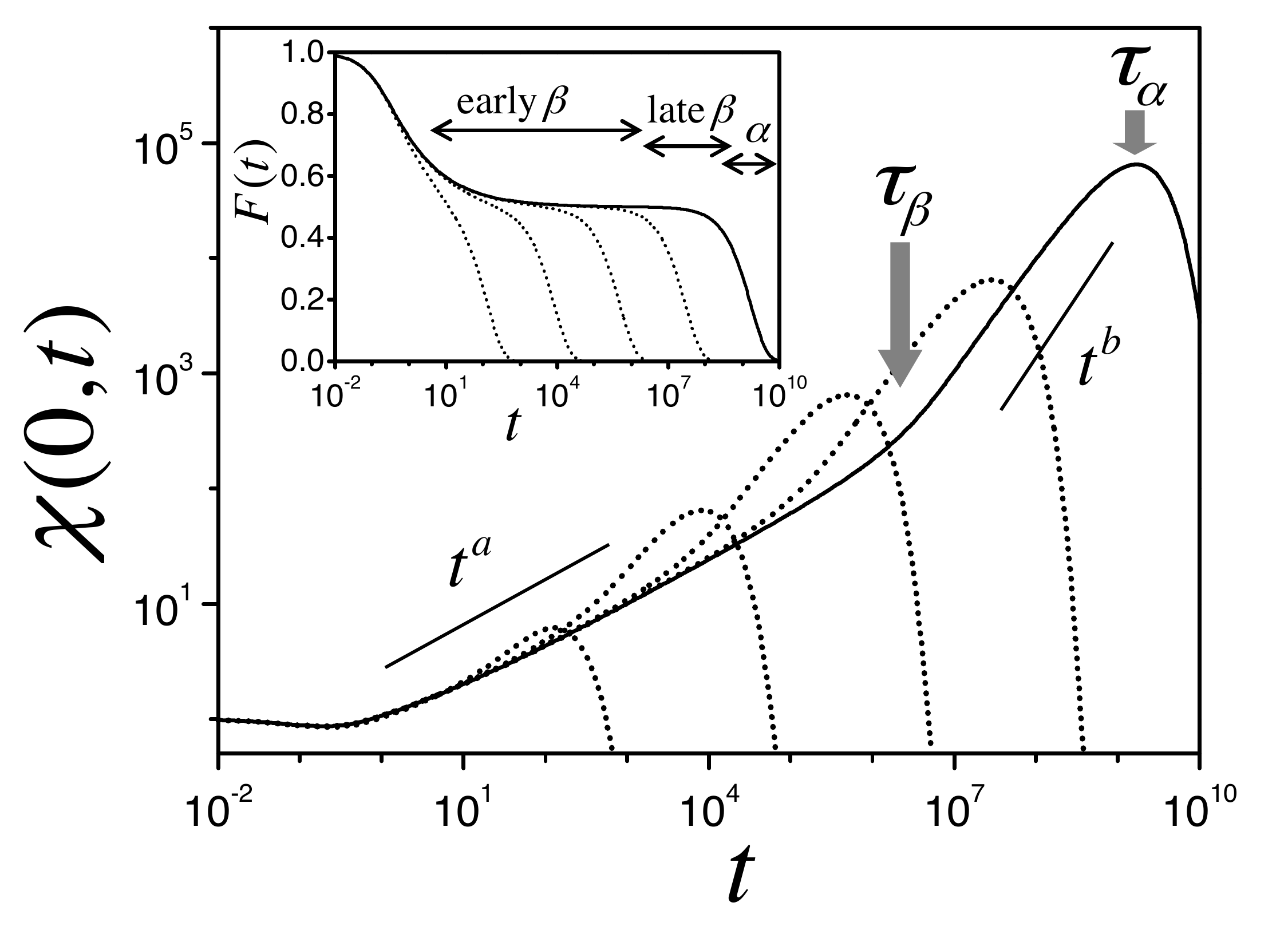} 
\caption{
$\chi(q=0,t)$ for various values of $\varepsilon$($=10^{-1} \sim 10^{-5}$).
Note that there are two algebraic growth regimes characterized by $t^a$ and
 $t^b$, respectively, below and above $\tau_{\beta}$. 
The peak height at $\tau_\alpha$ is scaled as 
$\chi^{\ast}\approx 1/\varepsilon$.
The inset reports the density correlation $F(t)$, corresponding to the main panel.
}
\vspace*{-0.8cm}
\label{fig-IMCT-q0}
\end{center}
\end{figure}

\begin{figure}[ht]
\begin{center}
\includegraphics[scale=0.38,angle=0]{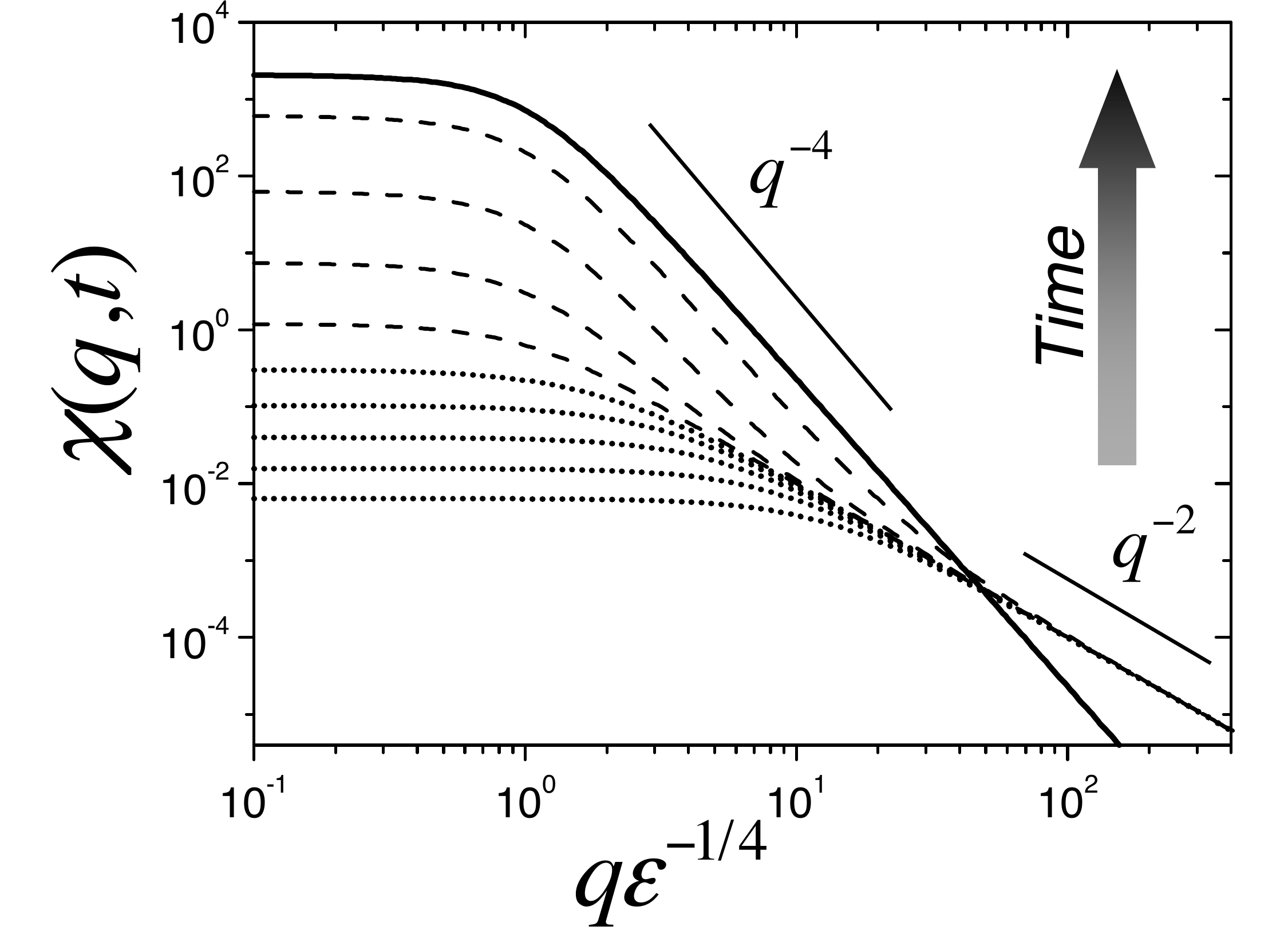} 
\caption{
Wave vector $q$ dependence of $\chi(q,t)$ for various $t$ from well below $\tau_\beta$
 up to $t\approx \tau_\alpha$.
$\varepsilon = |1- T/T_d|$ is fixed to $10^{-6}$. 
$q$ is scaled by $\varepsilon^{-1/4}$ ($\propto \xi(\tau_\alpha)$).
The dotted lines are Lorentzian functions capturing the early-$\beta$ regime. 
The solid line marks $t =\tau_\alpha$, where $1/q^4$ at large $q$. 
Between $\tau_\beta$ and $\tau_\alpha$ (the late-$\beta$ regime), one
observes a crossover from one regime to the other.}
\vspace*{-0.8cm}
\label{fig-IMCT-q-dep}
\end{center}
\end{figure}

As discussed in Sec.~\ref{sec:chi4}, Eq.~\eqref{eq:eom} for the $p$-spin model is mathematically 
equivalent to the MCT equation for molecular fluids,
and the nonlinear susceptibility (either $\chi_{4}(t)$ or $\chi_{FP}(t)$) for the $p$-spin model 
captures features of the simulated $\chi_4(t)$ for the molecular glasses
qualitatively, even though the $p$-spin model is completely blind to the spatial information associated with particle dynamics.  
Due to this parallel, it is natural to expect that molecular MCT can be extended to the calculation of the non-linear susceptibility for molecular glasses. 
This extension has indeed been put forward in Refs.~\cite{BB2004,Biroli2006prl}.   
The key idea is to reformulate MCT in the presence of a spatially
modulated external field governed by a perturbed Hamiltonian
$U_{\mbox{\scriptsize ext}}({\bf q}) = \lambda \rho_{{\bf q}}$, where
${\bf q}$ is the wave vector associated with the spatial modulation by the external field. 
The derivative of the density-density correlation function
$F({\bf k}, {\bf k} + {\bf q}, t)$ with respect to the external perturbation is nothing but the three-point susceptibility 
$\chi_{\bf k}({\bf q}, t) \propto 
\delta F({\bf k}, {\bf k} + {\bf q}, t)/\delta U_{\mbox{\scriptsize ext}}({\bf q})$. It captures spatially-dependent dynamical correlations as it probes how much a perturbation at, say, the origin affects the dynamics at a distance $r$. If one considers the two-point dynamical correlation function as the order parameter of the glass transition, then $\chi_{\bf k}({\bf q})$ captures the critical behavior, just like the linear susceptibility does in standard second-order phase transitions. 
The function $\chi_{\bf k}({\bf q}, t)$ thus conveys the same information as the four-point correlation function $\chi_{4}(t)$, which measures the fluctuations of the dynamical overlap.  Importantly, this function inherently contains information on the length scale associated with the dynamical heterogeneity probed by the spatially modulated field at wave vector ${\bf q}$, information that is absent in the counterpart of the $p$-spin model. The resulting inhomogeneous MCT (IMCT) equation is
\begin{equation}
\begin{aligned}
&
\frac{\partial \chi_{\bf k}({\bf q},t)}{\partial t}
+ \mu_{\bf k} \chi_{\bf k}({\bf q},t)
+\int_{0}^{t}\!\!{d} \tau~ M_{\bf k}(t-\tau)
\frac{\partial \chi_{\bf k}({\bf q},\tau)}{\partial \tau}
\\
&
\hspace*{1.5cm}
+\int_{0}^{t}\!\!{d} \tau~ 
H_{\bf k}({\bf q},t-\tau)\frac{\partial
 F_{|{\bf k}+{\bf q}|}(\tau)}{\partial \tau}
={\cal S}_{\bf k}({\bf q},t),
\end{aligned}
\label{eq:imct}
\end{equation}
where $\mu_{\bf k}$ is a diffusion coefficient, 
${\cal S}_{\bf k}({\bf q},t)$ is an inhomogeneous source term that does not affect the critical behavior,
$M_{\bf k}(t)$ is the memory kernel of the conventional MCT equation, and
$H_{\bf k}({\bf q},t)$ is given by
\begin{equation}
\begin{aligned}
&
H_{\bf k}({\bf q},t)
=
\frac{2|{\bf k}|}{|{\bf k}+{\bf q}|}
\int\!\!{d}{\bf k}^{\prime} V_{\bf k}({\bf k}^{\prime},{\bf k}-{\bf k}^{\prime})
V_{{\bf k}+{\bf q}}({\bf k}-{\bf k}^{\prime},{\bf q}+{\bf
 k}^{\prime})\chi_{{\bf k}'}({\bf q},t)F_{|{\bf k}-{\bf k}'|}(t)
\end{aligned}
\label{eq:imct-memory}
\end{equation}
with the vertex function of  the conventional MCT $V_{\bf q}({\bf k},{\bf k}^{\prime})$.
Equation~\eqref{eq:imct} has been analyzed theoretically and solved numerically~\cite{Biroli2006prl}. 
The overall behavior of the solution of Eq.~\eqref{eq:imct} can be inferred from the
scaling behavior of the MCT equation near the dynamical transition point
$T_d$. 
The $\beta$ regime, i.e, the time window close to $\tau_\beta \equiv |T-T_d|^{-1/2a}$ (where $a$ is a specified MCT exponent), corresponds to particles largely staying within 
the cages formed by their neighbors. We then find
\begin{equation}
\chi_{\bf k}({\bf q},t)=
\cfrac{C_{\bf k}}{\sqrt{\varepsilon} + \Gamma q^2}
g_{\beta}(\Gamma q^2/\sqrt{\varepsilon}, t/\tau_\beta)
\label{eq:qscaling-beta}
\end{equation}
with $\Gamma$ a constant and $C_k$ a weak function of $k= |\vec k|$.  
$\varepsilon \equiv |1-T/T_d|$ is the scaled distance from the dynamical transition point. 
$g_\beta(x,y)$ is a scaling function which ensures the early-$\beta$ relaxation
$g(0,y) \sim y^{a}$ at $y \rightarrow 0$ and the late-$\beta$ relaxation
$\sim y^{b}$ (where $b$ is another MCT exponent)
at $y \rightarrow \infty$, which seamlessly converges to
the scaling behavior in the $\alpha$-relaxation regime. 
By contrast, the scaling in the $\alpha$ regime, in which particles escape their cages, is characterized by
$\tau_\alpha = \varepsilon^{-\gamma}$ (with $\gamma = (1/a+1/b)$), which is given by  
\begin{equation}
\chi_{\bf k}({\bf q},t)=
\cfrac{1}{\sqrt{\varepsilon}(\sqrt{\varepsilon} + \Gamma q^2)}
f(\Gamma q^2/\sqrt{\varepsilon})g_{\alpha,k}(t/\tau_\alpha),
\label{eq:qscaling-alpha}
\end{equation}
where the scaling function $f(x)$ behaves as $\sim 1/x$ 
at $x \gg 1$, from the condition that the length scale should become
independent of $\varepsilon$.
Both scalings, Eqs.~\eqref{eq:qscaling-beta} and \eqref{eq:qscaling-alpha}, 
assert that the dynamical length scale should diverge as $\xi \propto \varepsilon^{-\nu}$, with $\nu=1/4$ rather than $1/2$ as for a Landau theory. 
These asymptotic scalings can be checked by the full wave vector dependent solution of the IMCT equation. 
However, integrating Eq.~\eqref{eq:imct} numerically is a formidable task 
due to the coupling of the two wave vectors $k$ and $q$. 
From the analogy that the schematic MCT (Eq.~\eqref{eq:eom} for the $p$-spin model)
captures the main features of the dynamical behavior of the full $k$-dependent MCT equation, 
we simplify the IMCT equation by removing one of the wave vectors, $k$, which monitors only the
static microscopic length of the order of the molecular size.
This schematic IMCT equation is numerically integrated and the results 
for $q=0$ are shown in Fig.~\ref{fig-IMCT-q0}; the $q$ dependence of various time
regimes are shown in Fig.~\ref{fig-IMCT-q-dep}. 
The behavior of $\chi(q=0,t)$ is similar to that of the $p$-spin model in
Fig.~\ref{fig3}, characterized by the two scalings $t^{a}$ and $t^{b}$ for 
the early-$\beta$ ($t <\tau_\beta$) and the late-$\beta$ regimes
($\tau_\beta < t < \tau_\alpha$), respectively, followed by a growing
peak $\chi(q=0, t=\tau_\alpha) \approx \varepsilon^{-1}$ at $t \approx \tau_\alpha$. 
The $q$ dependence of $\chi(q,t)$ is interesting as it demonstrates
the rich hierarchical growth of the fluctuations. 
First, in the early-$\beta$ regime ($t < \tau_\beta$),  $\chi(q,t) \approx {\xi^2(t)}/({1 + q^2 \xi^2(t)})$
is characterized by an Ornstein-Zernike form and the dynamical correlation
length grows with time as $\xi \propto t^{a/2}$, algebraically
characterized by the MCT exponent $a$. 
As the system enters the late-$\beta$ regime, $\xi$ stops growing and is given by
$\varepsilon^{-1/4}$ and, concomitantly, the shape of the spectrum develops a tail at large $q$,  which eventually behaves as $1/q^4$ at $\tau_\alpha$. 
These results imply that the morphology of the dynamical
heterogeneity changes non-trivially over time. As time progresses, 
fluctuations grow, but the shape of dynamically heterogeneous regions are more fractal at the early-$\beta$ regime and
then gradually fatten in the late-$\beta$ (and early-$\alpha$) regime, where dynamically heterogeneous regions become compact. 

Verifying these results in molecular glass formers by
simulations and experiments is a difficult task because MCT, and therefore IMCT, are mean-field descriptions
whose \emph{critical} behavior are washed out by thermal hopping and/or
facilitation dynamics in finite dimensional systems, especially in the $\alpha$ regime. 
Results of a recent simulation study, however, show nearly
quantitative agreement with IMCT in the $\beta$ regime. More specifically, these simulations show both the
algebraic growth of $\xi$ with $t$ and the power-law $\xi  \sim
\varepsilon^{-1/4}$~\cite{Tah2020prr}.  
This behavior is in harmony with the general notion that MCT works best in
the $\beta$ regime, where the collective dynamics of the unstable modes navigating
saddles in the energy landscape dominate. 
Although MCT and IMCT are only capable of describing the moderately supercooled regime
above $T_d$ or intermediate time scales shorter than the $\alpha$ regime, it is safe to claim that it is the sole first-principles theory that describes the hierarchically rich dynamics over several
decades without a single fitting parameter. 
A generalization of IMCT to the higher-order glass singularities that are found, for instance,
deep in the repulsive-attractive glass-forming regime of certain liquids has also been proposed~\cite{Nandi2014prl}.

\section{From mean-field theory to finite dimensions}
 In this section, we first present the building blocks of the theory of dynamical fluctuations going from mean-field to finite-dimensions. We then present the numerical results obtained in a three dimensional model of glassy liquids 
where non-mean field effects are suppressed.
\subsection{Breaking up $\chi_4$: different kinds of dynamical fluctuations}
\label{sec:squaring}
Section~\ref{sec:pspin} focused on dynamical susceptibilities rather than on dynamical correlations. In particular, all $\chi$'s ( e.g., $\chi_T$) were obtained as the response of a suitable correlation function to changing a control parameter. These susceptibilities were originally thought to scale the same way as $\chi_4$ does. Only later, when considering finite dimensional fluctuations around MCT, was it realized that this is not the case: a \emph{squaring} effect emerges in $\chi_4$. Its physical origin can be understood by splitting the fluctuations of the dynamical correlation $C(t,0)$ into two parts:
\begin{equation}
    \chi_4(t)=\langle\left(C(t,0)-\langle C(t,0) \rangle_T \right)^2\rangle_{T,IC}+\langle\left(\langle C(t,0) \rangle_T-\langle C(t,0) \rangle_{T,IC} \right)^2\rangle_{T,IC}.
\end{equation}
The first contribution describes the fluctuations at fixed initial condition (due to thermal noise), while the second describes the fluctuations due to the initial conditions. The latter dominates. This was first realized in terms of a liquid-state diagrammatic field theory in \cite{Berthier2007jcp1,Berthier2007jcp2}, and explicitly demonstrated and further studied in molecular dynamics simulations in \cite{berthier2007structure}. A full understanding was reached after the work \cite{franz2011field,franz2012quantitative}, which used replica field theory to study dynamical fluctuations in the $\beta$ regime (first using the analogy between disordered systems and glasses in \cite{franz2011field} and then later directly in glassy liquids \cite{franz2012quantitative}).

The authors of Ref.~\cite{franz2011field} showed that the statistics of the overlap fluctuations at the plateau, i.e. in the $\beta$ regime, are the same as those associated with the spinodal of the random field Ising model (RFIM) \cite{nattermann1998theory}. In this mapping, the spinodal is the counterpart of the MCT transition, whereas the disorder is the analog of the initial metastable state, in which the system resides as set by the initial condition. 

In terms of scaling, this set of results implies that 
\begin{equation}\label{eq:square}
    \chi_4\sim \chi_T^2,
\end{equation}
where, as mentioned above, $\chi_{T}=\frac{dC(t)}{dT}$. In other words, the dynamical fluctuations are proportional to the square of the dynamical susceptibility. This scaling can be understood by noticing that some small fluctuations in the initial conditions (the fluctuations related to observables that relax slowly) lead to giant fluctuations of $C(t,0)$. This amplification is governed by the dynamical susceptibilities. For instance, a metastable state with a free energy which is slightly lower than the average has a much longer relaxation time, and hence a correlation function with a much longer plateau. Therefore, there is a component of the fluctuations that can be roughly written as $\delta C(t,0)=\chi_T \frac{dT}{df}\delta f$. It is the square of this contribution which leads to the scaling relation given by Eq.~\eqref{eq:square}. Note that this is the same type of argument used to relate connected and disconnected susceptibilities in the RFIM \cite{nattermann1998theory}. 

The mapping between the field theory of the overlap fluctuations and the spinodal of the RFIM has played a very important role in firmly establishing the importance of self-induced disorder fluctuations 
\cite{biroli2018random1,biroli2018random2},  
already highlighted in \cite{stevenson2008constructing}, and in opening the way to a finite dimensional analysis of the MCT transition. This analysis is very intricate, as the spinodal of the RFIM cannot be studied perturbatively \cite{nandi2016spinodals,rizzo2016dynamical}. We shall return to this point in the conclusions. 

\subsection{Gaussian core model}
In finite dimensions, numerical verification of the mean-field description given by MCT has
never been satisfactory. Thermally activated dynamics 
and other mechanisms such as dynamical facilitation, which are generically local, 
obfuscate the mean-field physics as temperature is lowered. 
The window over which the MCT power-law divergence of the relaxation
quantitatively describes relaxation behavior is narrow in realistic simulations of three-dimensional liquids. 
Likewise, the agreement between the growing dynamical heterogeneity predicted by IMCT and that of molecular dynamics is largely qualitative. A realistic particle-based model system in finite dimensions, which compellingly verifies the behaviors predicted by MCT and IMCT, has therefore long
been sought. 
The Gaussian core model (GCM) is such a model~\cite{Ikeda2011prl}. 
This system describes a monatomic fluid whose interaction is pair-wise and Gaussian. 
Contrary to other monatomic glass models, the nucleation rate of the GCM is
extremely small at high densities, and thus the GCM does not crystallize even without size polydispersity. 
Ikeda and coworkers have found that the slow dynamics of the supercooled GCM is
described unprecedentedly well by MCT and IMCT~\cite{Ikeda2011prl,Coslovich2016pre}. 
In particular, the dynamical transition temperature $T_d$ is closer to
that predicted by MCT and the window of power-law scaling is wider than for any other simulated glass former thus far. In addition, the dynamics are spatially very uniform. 
For typical glass formers, such as Lennard-Jones binary mixtures, the
distribution of particle displacements exhibits clear bimodal structure, thus clearly separating particles into fast and slow
groups. This behavior has
been believed to be one of the signatures of heterogeneous dynamics. 
However, the distribution of particle displacements in the GCM is singly-peaked and nearly Gaussian, even in the deeply supercooled regime. 
More surprising is the fact that dynamical heterogeneity is extremely strong in the GCM despite the
near-Gaussian statistics of displacements~\cite{Coslovich2016pre}. 
\begin{figure}[ht]
\centerline{\includegraphics[width=10cm]{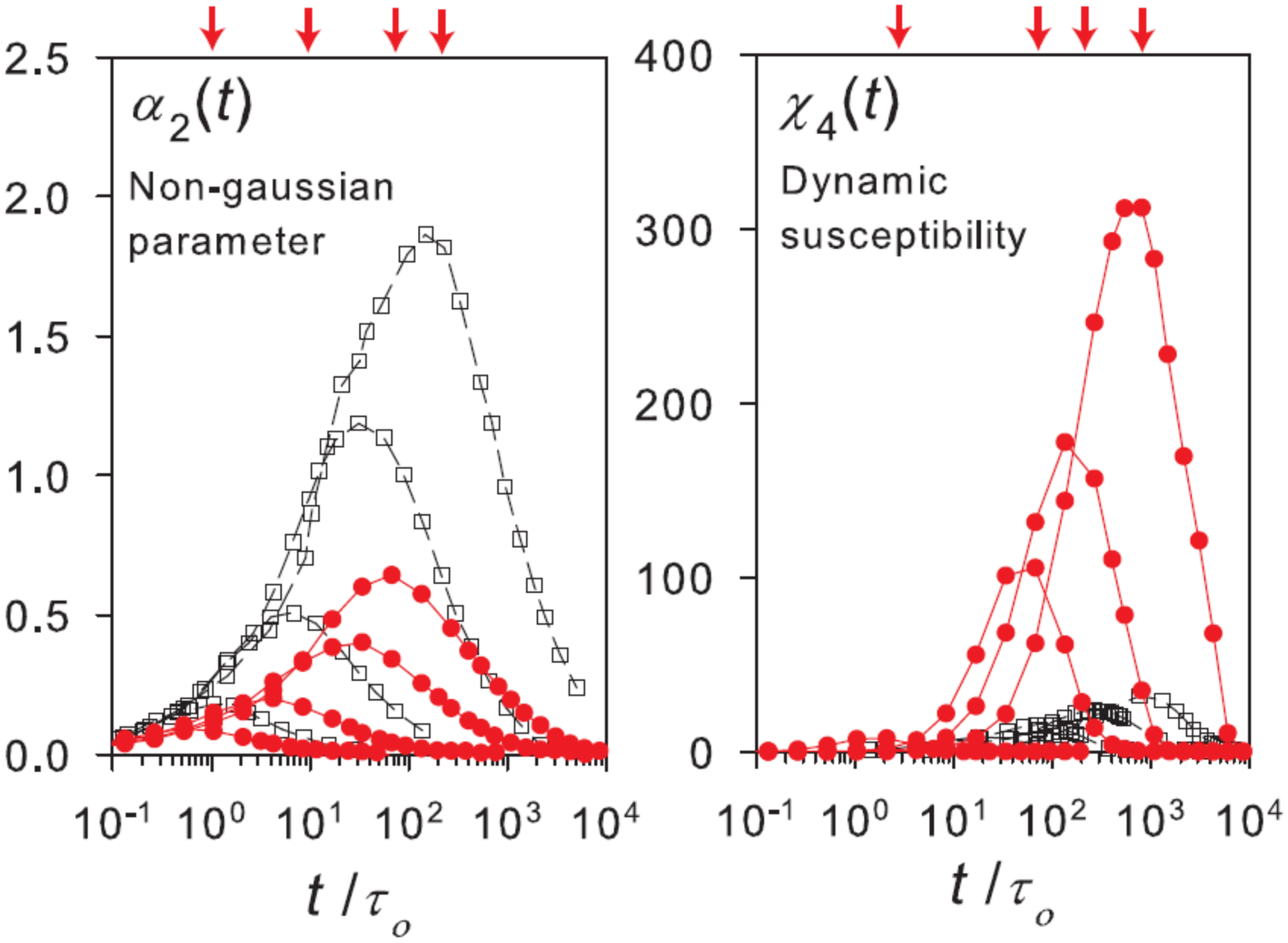}}
\caption{The non-Gaussian parameter $\alpha_2(t)$ (left) and the non-linear
 susceptibility $\chi_4(t)$ (right) of the GCM (circles) and the KA
 mixture (empty squares) for several temperatures near their respective dynamic
 transition points.  The time is scaled by the relaxation time at their respective
 onset temperatures, $\tau_o$. Taken from Ref.~\cite{Coslovich2016pre}.}
\label{fig-Coslovich2016pre}
\end{figure}
Figure~\ref{fig-Coslovich2016pre} shows the non-Gaussian parameter
$\alpha_2(t)$ and the nonlinear susceptibilities $\chi_4(t)$ for the GCM
and the Kob-Andersen binary Lennard-Jones (KA) mixture, respectively. 
$\alpha_2(t)$, which monitors the deviation of the distribution of particle 
displacements from a Gaussian, is markedly smaller in the GCM than in
the KA mixture at similar degrees of supercooling. 
For $\chi_4(t)$, the situation is reversed. The peak
heights for the GCM are larger by more than an order of magnitude
than for the KA mixture. 
Furthermore, the temperature dependence of the peak height of
$\chi_4(t)$ for the GCM is well fitted by 
\begin{equation}
 \chi_4(t=\tau_\alpha) \sim  |T-T_d|^{-2},
\end{equation}
up to very close to $T_d$, which is in excellent agreement with the IMCT
and mean-field predictions (taking into account the \emph{squaring} effect discussed in Sec.~\ref{sec:squaring},
Eq.~\eqref{eq:square}). 
This set of agreements is not accidental. The analysis of the
modes associated with saddle points on the energy landscape shows that
the participation ratio of these modes are 
more extended than in typical glass formers such as the KA mixture, strongly
indicating that unstable modes are delocalized. 
This is exactly the scenario encoded in MCT, namely
that the dynamical transition of MCT is a geometrical transition with
a diverging cooperative length scale over unstable saddles, the number
of which vanishes as $T_d$ is approached from above~\cite{Biroli2012wolynesbook,Coslovich2019scipost}. 
In addition, the average activation energy in the GCM is far
higher than in the KA mixture, which implies that thermal activation processes are suppressed. 
It would be interesting to investigate the $q$ dependence of $\chi_4(q,t)$
for the GCM and verify the scaling behavior of IMCT. 
Such simulations, however, would be extremely expensive due to the long
cutoff length of the Gaussian interaction potential and the large overlaps with the nearest particles at high densities. 

\section{Conclusion and Perspectives}
Dynamical heterogeneity is a core aspect of the phenomenology of glass-forming liquids. The replica technique has played an important role in unveiling physical mechanisms behind the emergence of the phenomenon. There are several directions in which it may still prove instrumental.  

First, dynamical heterogeneity encompasses a set of phenomena that is broader than just spatial dynamical correlations \cite{berthier2011dynamical}. Can the replica technique be used to get insights into the full range of these phenomena? 
As a concrete example, we refer the reader to very recent work based on the large dimensional analysis of the glass transition\cite{biroli2022local}, which provides a new perspective on the emerging non-Gaussian behavior of local dynamical observables.  This and related lines of research are ripe for exploration with replica techniques.

The replica technique and its dynamical counterparts have provided a sound
platform to explain the development of the spatial and temporal behavior
of dynamical correlations approaching the MCT transition. However, in
real glass-forming liquids this is only the beginning of the story. In
fact, dynamical heterogeneity becomes stronger, more prominent and is
characterized by larger length scales approaching $T_g$, as remarkably
shown in very recent work \cite{scalliet2022thirty}. What can be said
about this behavior from replica theory? More specifically, as such behavior is clearly related to that of dynamical facilitation, does replica theory have anything to say about it? Approaches based on RFOT theory addressing some of these issues have been developed in \cite{bhattacharyya2008facilitation} (see also the recent discussion in \cite{biroli2022rfot}). An interesting and complementary research direction would consider the non-perturbative corrections based on the mapping to the RFIM, as discussed above.  These corrections are different from the non-perturbative ones associated to the growth of a static length-scale connected to growing amorphous order. They are instead related to an avalanche effect in which a region more prone to rearrange induces cascades of rearrangements nearby, as it happens for the spinodal of the RFIM \cite{nandi2016spinodals}. What remains to be done is to translate these ideas, based on the theory of static fluctuations of the overlap, in a fully fledged dynamical theory. This program is certainly not easy. Except close to $T_{d}$ and in the $\beta$ regime \cite{rizzo2016dynamical}, this is a fully open problem. It is nevertheless one worth trying, as it would offer a detailed, replica-based theory of dynamical facilitation. 
\\\\

\subsection*{Acknowledgments}  We want to deeply thank all our collaborators on the topics covered in this chapter, in particular L. Berthier, J.-P. Bouchaud, 
D. Coslovich, O. Dauchot, A. Ikeda,
 F. Ladieu, G. Tarjus, M. Tarzia for many discussions on all these issues over the years. GB, DRR, and KM are members of the Simons Foundation "Cracking the Glass Problem" collaboration. We would like to thank all of our colleagues in this collaboration for many years of stimulating discussions. GB and DRR are partially supported by the Simons Foundation (GB-Grant No. 454935, DRR-Grant No. 454951).  KM is financially supported by KAKENHI 20H00128. 

\bibliographystyle{ws-book-har}    
\bibliography{bib_DH}
\end{document}